\documentstyle[12pt]{article}

\begin{document}

\title{\Large {\bf  The Bekenstein-Hawking Entropy of 
                    Higher-Dimensional Rotating Black Holes } }

\author{{{\large Zheng Ze Ma}} \thanks{E-mail: 
           z.z.ma@seu.edu.cn} 
  \\  \\ {\normalsize {\sl Department of Physics, Southeast 
                 University, Nanjing, 210096, P. R. China } } }

\date{}

\maketitle

\begin{abstract}

\indent

  A black hole can be regarded as a thermodynamic system 
described by a grand canonical ensemble. In this paper, we study 
the Bekenstein-Hawking entropy of higher-dimensional rotating 
black holes using the Euclidean path-integral method of Gibbons 
and Hawking. We give a general proof demonstrating that ignoring 
quantum corrections, the Bekenstein-Hawking entropy is equal to 
one-fourth of its horizon area for general higher-dimensional 
rotating black holes.

\end{abstract}

\section{Introduction}

\indent

  The Bekenstein-Hawking entropy of a black hole, not considering 
quantum corrections, is equal to one-fourth of its horizon 
area.$^{1),2)}$ This conclusion can be obtained using many different 
approaches. A classical approach to derive the entropy of a black hole 
was formulated by Gibbons and Hawking in terms of the Euclidean 
path-integral and grand canonical ensemble methods.$^{3),4)}$ In this 
paper, we study the entropy of higher-dimensional rotating black holes 
following the method of Gibbons and Hawking. To make this work 
self-contained, we first review the method of Gibbons and Hawking 
for determining the entropy of a black hole.$^{3),4)}$

  A black hole can be regarded as a thermodynamic system described by 
a grand canonical ensemble, because it possesses an explicit 
temperature and is in a state of thermal equilibrium with respect to 
radiation. It exchanges particles and energy with the surrounding  
spacetime. We can write down its grand partition function $Z$, 
thermodynamic potential $W$ and entropy $S$: 
$$
  Z=\exp[-W]={\rm Tr} \exp[-(\beta \hat{H}-\mu_{i}\hat{C_{i}})] ~,
  \eqno{(1)}  $$
$$
  W=E-TS-\mu_{i}C_{i} ~,
  \eqno{(2)}  $$
$$
  S=\beta(E-\mu_{i}C_{i})+\ln Z ~. 
  \eqno{(3)}  $$
Here $\frac{1}{\beta}$ is the temperature and $\mu_{i}$ represents the 
chemical potentials. The partition function (1) can also be written in 
the Euclidean path integral form. This is given by 
$$
  Z=\int D[g,\phi]\exp(-I_{E}[g,\phi]) ~,
  \eqno{(4)}  $$
where $\phi$ represents matter fields, including the electromagnetic 
fields of charged black holes. A Wick rotation has been performed in 
order to realize the Euclidean form in (4). We can compute $Z$ 
perturbatively, and to first order, we obtain  
$$
  Z=\exp[-I_{E}^{\infty}] ~, 
  \eqno{(5)}  $$
where
$$
  I_{E}^{\infty}=\frac{1}{16\pi}\int_{M}[(R-2\Lambda)+
        {\cal L}_{{\rm matter}})]+\frac{1}{8\pi}\int_{\partial M}[K]
  \eqno{(6)}  $$
is the on-shell Euclidean action, and $[K]=K-K_{0}$ is the difference 
between the extrinsic curvature of the spacetime manifold and that of a 
reference background spacetime. The upper index $\infty$ of (6) means 
that the boundary of the spacetime manifold lies only at $r=\infty$, 
i.e., in (6) $\partial M$ lies only at $r=\infty$. The reason for this 
is that we are considering the black hole as a thermal equilibrium system. 
Thus it is reasonable that we take the metric of the spacetime manifold 
in the form of the largest extension, as in the case of the Kruskal 
metric of a Schwarzschild black hole, because the singularity on the 
horizon is only apparent and can be moved away.$^{5),6)}$ Also $I_{E}$ 
is expected to be invariant under a general coordinate transformation. 
The above arguements also hold for the metric of a higher-dimensional 
rotating black hole. For the metric of a higher-dimensional rotating 
black hole, the above consideration is also tenable.

  Next, we must treat the term $\beta(E-\mu_{i}C_{i})$ in (3). For 
this purpose, we consider the quantum transition amplitude between two 
space-like hyperspaces in a black hole's spacetime manifold in the 
Euclidean time formalism. This is given by 
$$
  \langle\tau_{1}\vert \tau_{2}\rangle=\langle\tau_{1}\vert 
     e^{-(\tau_{2}-\tau_{1})\hat{H}}\vert\tau_{1}\rangle ~.
  \eqno{(7)}  $$
Under the condition that the fluctuations of energy are relatively 
small, i.e., $\frac{\langle E\rangle^{2}-\langle E^{2}\rangle}
{\langle E\rangle^{2}}\ll 1$, we can expand (7) while ignoring the 
fluctuations. This yields  
$$
  \langle\tau_{1}\vert \tau_{2}\rangle= e^{-(\tau_{2}-\tau_{1})E} ~.
  \eqno{(8)}  $$
For example, for a Schwarzschild black hole, we have 
$\langle E\rangle^{2}=M^{2}$, 
$\langle E\rangle^{2}-\langle E^{2}\rangle=\frac{M^{2}_{P}}{8\pi}$, 
and have the above condition is satisfied if $M\gg M_{P}$. As pointed 
out by Kallosh et al.,$^{7)}$ (8) can be generalized to the case in 
which there exist multiple conserved charges $C_{i}$. To introduce 
Lagrange multipliers $\mu_{i}$ and to 
consider constrained imaginary time evolution so that only metrics 
with designated charges are considered in the path integral, we 
obtain
$$
  \langle\tau_{1}\vert \tau_{2}\rangle= e^{-(\tau_{2}-\tau_{1})
      (E-\mu_{i}C_{i})} ~.
  \eqno{(9)}  $$

  On the other hand, the quantum transition amplitude 
$\langle\tau_{1}\vert \tau_{2}\rangle$ can be obtained using the 
Feynman path integral formulation as  
$$
  \langle\tau_{1}\vert \tau_{2}\rangle=\int~ D[g,\phi]e^{-I[g,\phi]} ~.
  \eqno{(10)}  $$
To first-order, this gives 
$$
  \langle\tau_{1}\vert \tau_{2}\rangle=e^{-I^{\infty}_{E,h}} ~,
  \eqno{(11)}  $$
where again $I^{\infty}_{E,h}$ is given by (6). Here the subscript 
$h$ on the Euclidean action means that the black hole's horizon is 
considered as another spacetime boundary. We include this boundary 
because we study the quantum transition amplitude in the black hole's 
spacetime manifold. No physical information can escape from the horizon 
of a black hole. Therefore we need to take the horizon as another 
spacetime boundary. Thus the black hole's horizon also contributes to 
the Euclidean action integral of (6) in (11). Comparing (9) and (11), 
and fixing
$$
  \tau_{2}-\tau_{1}=\beta ~,
  \eqno{(12)}  $$
we obtain
$$
  \beta(E-\mu_{i}C_{i})=I^{\infty}_{E,h} ~.
  \eqno{(13)}  $$
This relation results from the fact that $\beta$ is the period of the 
imaginary time for the black hole's spacetime. Inserting (5), (6) and 
(13) into (3), we obtain 
$$
  S=\frac{1}{8\pi} \left (\int^{\infty}_{h}
      [K]-\int^{\infty}[K] \right )=-\frac{1}{8\pi}\int_{h}[K] ~. 
  \eqno{(14)}  $$
Thus we arrive at the conclusion of Gibbons and Hawking: A black 
hole's entropy, not considering quantum corrections, is determined by 
the gravitational surface term. This conclusion is valid for 
spherically symmetric black holes, as well as for charged and rotating 
black holes. The differences among the entropies of different types of 
black holes result only from their different conserved charges $C_{i}$, 
because the matter fields parts are canceled in the above derivation.

\section{The Bekenstein-Hawking entropy of 
                    higher-dimensional rotating black holes}

\indent

  Using certain surface terms of the gravitational action, one can 
derive the relation $S=\frac{1}{4}A$ for a black hole's entropy. The 
derivation of this relation is given in Refs. 3), 7) and 8) for certain 
spherically symmetric black holes. We seek to derive the relation 
$S=\frac{1}{4}A$ for higher-dimensional rotating black holes in this 
paper. For this purpose, we need to derive the explicit form of the 
gravitational surface term $K$.

  There are many definitions of the gravitational surface term in the 
literature (e.g., see Refs. 3), 4) and 7)-10)). Usually, the 
surface term $K$ can be defined as the trace of the second fundamental 
form on the horizon. However, different forms of the gravitational 
surface terms are equivalent for the purpose of determining a black 
hole's entropy according to the formula (14). This is because the 
gravitational surface terms are total derivatives decomposed from the 
Einstein gravitational action. For different choices of the surface 
term, their differences are also total derivatives. Clearly, the field 
equations are not changed when a total derivative is added to the 
gravitational action. Therefore, the metric of a black hole is not 
changed when a total derivative is added to the gravitational action. 
Hence, different choices of the surface term yield identical forms of 
the black hole entropy formula (14). We find that the surface term of 
Landau and Lifshitz$^{10)}$ is most convenient for deriving the relation 
$S=\frac{1}{4}A$ for higher-dimensional rotating black holes, as in the 
case of spherically symmetric black holes.$^{7),8)}$

  According to Landau and Lifshitz,$^{10)}$ the gravitational action can 
be decomposed in two parts
$$
  \sqrt{-g}R=\sqrt{-g}G+\partial_{\mu}(\sqrt{-g}\omega^{\mu}) ~,
  \eqno{(15)}  $$ 
where the first term contains no second derivatives. Omitting the total 
derivative $\partial_{\mu}(\sqrt{-g}\partial_{\nu}g^{\mu\nu})$ in the 
action, $\omega^{\mu}$ is obtained as 
$$
  \omega^{\mu}=-\frac{2}{\sqrt{-g}}\left( \frac{\partial\sqrt{-g}}
                   {\partial x^{\nu}} \right)
       g^{\mu\nu}-\frac{\partial g^{\nu\mu}}{\partial x^{\nu}} ~.
  \eqno{(16)}  $$
Therefore, the surface term is given by
$$
  K=\frac{1}{2}\omega^{\mu}n_{\mu} ~,
  \eqno{(17)}  $$
where $n_{\mu}$ is the space-like outward-pointing unit normal vector 
on the horizon. Usually, in the case of a charged black hole, there 
also exist certain surface terms that are related to the matter 
fields$^{11),12)}$ as 
$$
  K_{\mbox{{\tiny matter}}}=\frac{1}{4\pi}\int_{h}d^{D-1}x\sqrt{g_{D-1}}
        n_{\mu}F^{\mu\nu}A_{\nu} ~,
  \eqno{(18)}  $$
where $D$ is the dimension of the spacetime manifold. However, in the 
derivation of (14), the contributions from the matter fields are 
canceled. Therefore we need not to consider the surface terms that 
related to different matter fields in the entropy formula (14). 
However, these terms may have an effect in the analysis of quantum 
corrections to the black hole's entropy.

  Now we rewrite (14) in the explicit form
$$
  S=-\frac{1}{8\pi}\int_{h}d^{D-1}x\sqrt{g_{D-1}}
      (K-K_{0}) ~,
  \eqno{(19)}  $$
where $D$ is the dimension of the spacetime manifold, $g_{D-1}$ is the 
determinant of the metric on the horizon, and $K_{0}$ comes from the 
background Minkowski spacetime in which the black hole is embedded. The 
general solution for a higher-dimensional rotating black hole has been 
obtained by Myers and Perry.$^{13)}$ Here we adopt the parameterized 
form derived by Cveti\u{c} and Youm$^{14)}$ for the purpose of deriving 
the relation $S=\frac{1}{4}A$. With this form, the metric for a 
higher-dimensional rotating black hole is given by
\begin{eqnarray*}
~~~~~~~~~~
  ds^{2} & = & g_{\tau\tau}d\tau^{2}+g_{rr}dr^{2}+
         g_{\theta\theta}d\theta^{2}+
         g_{\psi_{i}\psi_{i}}d\psi_{i}^{2}+
        2g_{\theta\psi_{i}}d\theta d\psi_{i}+
        \sum_{i<j}2g_{\psi_{i}\psi_{j}}d\psi_{i}d\psi_{j}   \\
         & ~ & +g_{\phi_{i}\phi_{i}}d\phi_{i}^{2}+
        \sum_{i<j}2g_{\phi_{i}\phi_{j}}d\phi_{i}d\phi_{j}+ 
        2g_{\tau\phi_{i}}d\tau d\phi_{i} ~.
~~~~~~~~~~~~~~~~~~~~~~~~~~~~~~~~~~~~~~ (20) 
\end{eqnarray*}
Here we have written the metric in Euclidean form, because 
formula (14) is derived for the Euclidean form of the black hole 
metric. The Euclidean form of the metric can be obtained from the 
corresponding Lorentzian metric by rotating the time axis and the 
parameters.$^{15),16)}$ In (20), $i$ and $j$ on $\phi$ run from 
$1$ to $[\frac{D-1}{2}]$. For the even-dimensional case, $i$ and $j$ 
on $\psi$ run from $1$ to $[\frac{D-3}{2}]$. For the odd-dimensional 
case, $i$ and $j$ on $\psi$ run from $1$ to $[\frac{D-5}{2}]$. 
Obviously, the black hole rotates in each of the $\phi_{i}$ 
directions, and there are $[\frac{D-1}{2}]$ angular momentum 
components. The area of the horizon is given by the integral 
$$
  A=\int d\theta \prod d\psi_{i} 
             \prod d\phi_{j}
            \Big[\sqrt{\det g_{ab}}\Big]\Big \vert_{r=r_{h}} ~.
  \eqno{(21)}  $$
Here we use $r_{h}$ to denote the radius of the horizon. Also, 
we use $g_{ab}$ to denote the metrics of the $d\theta^{2}$, 
$d\psi_{i}^{2}$, $d\theta d\psi_{i}$, $d\psi_{i}d\psi_{j}$, 
$d\phi_{i}^{2}$ and $d\phi_{i}d\phi_{j}$ terms of the metric given 
in (20). We regard the horizon to be the outer horizon of a rotating 
black hole. In addition, we 
consider non-extremal black holes in this paper. The entropy of an 
extremal black hole is zero, according to Refs. 8), 17) and 18). 
This can also be seen from (19). Because the radius $r_{h}$ of the 
horizon lies at the space-like infinity for an extremal black hole, 
we obtain $K-K_{0}=0$ in (19).

  We denote the angular velocity of the horizon corresponding to 
the coordinate $\phi_{i}$ as $\Omega_{i}$. In order to simplify 
the derivation, we carry out the coordinate transformation 
$$
  \phi_{i}\rightarrow\phi_{i}^{\prime}=\phi_{i}-\Omega_{i} \tau~, 
  \eqno{(22)}  $$
which means that the observer is co-rotates with the outer horizon. 
The metric (20) now becomes 
\begin{eqnarray*}
~~~~~~~~~~
  ds^{2} & = & G_{\tau\tau}d\tau^{2}+g_{rr}dr^{2}+
           g_{\theta\theta}d\theta^{2}+
         g_{\psi_{i}\psi_{i}}d\psi_{i}^{2}+
        2g_{\theta\psi_{i}}d\theta d\psi_{i}+
        \sum_{i<j}2g_{\psi_{i}\psi_{j}}d\psi_{i}d\psi_{j}       \\
         & ~ & +g_{\phi_{i}\phi_{i}}d\phi_{i}^{\prime~2}+
        \sum_{i<j}2g_{\phi_{i}\phi_{j}}
         d\phi_{i}^{\prime}d\phi_{j}^{\prime}+ 
        2g_{\tau\phi_{i}^{\prime}}d\tau d\phi_{i}^{\prime} ~,
~~~~~~~~~~~~~~~~~~~~~~~~~~~~~~~~~~~~ (23) 
\end{eqnarray*}
where
$$
    G_{\tau\tau}=g_{\tau\tau}+g_{\phi_{i}\phi_{i}}\Omega_{i}^{2}+
         \sum_{i<j}2g_{\phi_{i}\phi_{j}}\Omega_{i}\Omega_{j}+
         2g_{\tau\phi_{i}}\Omega_{i} ~,          $$      
$$
  g_{\tau\phi_{i}^{\prime}}=\Omega_{j}g_{\phi_{i}\phi_{j}}
                  +g_{\tau\phi_{i}} ~.
  \eqno{(24)}  $$
In the reference system co-rotating with the horizon, the metric on the
horizon is static. Therefore, $g_{\tau\phi_{i}^{\prime}}$ is zero on the 
horizon, and the angular velocities at the horizon satisfy
$$
  g_{\tau\phi_{i}^{\prime}}\vert_{r=r_{h}}=0 ~,
  \eqno{(25)}  $$
under the condition $\det g_{\phi_{i}\phi_{j}}\neq 0$. Then according 
to (A$~\cdot~$5), we have 
$$
  G_{\tau\tau}\vert_{r=r_{h}}=0 ~.
  \eqno{(26)}  $$
The metrics $g_{rr}$, $g_{\theta\theta}$, $g_{\psi_{i}\psi_{i}}$, 
$g_{\theta\psi_{i}}$, $g_{\psi_{i}\psi_{j}}$, $g_{\phi_{i}\phi_{i}}$ 
and $g_{\phi_{i}\phi_{j}}$ are unchanged under the coordinate 
transformation (22). For the metric (20), the quantity 
$g^{rr}=1/g_{rr}$ is also unchanged. The horizons of the metrics (20) 
and (23) are determined by the relation $g^{rr}=0$. Therefore, the 
coordinate transformation (22) does not change the location of the 
horizon.

  Because $g_{\tau\phi_{i}^{\prime}}$ is zero on the horizon, the 
determinant of the metric (23) on the horizon is 
$$
  g_{D-1}=G_{\tau\tau}\det g_{ab} ~,
  \eqno{(27)}  $$
where $g_{ab}$ denotes the metrics of the $d\theta^{2}$, 
$d\psi_{i}^{2}$, $d\theta d\psi_{i}$, $d\psi_{i}d\psi_{j}$, 
$d\phi_{i}^{\prime~2}$ and $d\phi_{i}^{\prime}d\phi_{j}^{\prime}$ 
terms in the metric (23). From (19), the entropy of the metric (23) 
is given by 
$$
  S=-\frac{1}{8\pi}\int_{0}^{\beta}d\tau
    \int d\theta \prod d\psi_{i} \prod d\phi_{j}^{\prime}~
    \frac{1}{2}[\omega^{\mu}n_{\mu}\sqrt{g_{D-1}}]
    \big\vert_{r=r_{h}} ~,
  \eqno{(28)}  $$
where $\beta=2\pi/\kappa$ is the inverse Hawking temperature, and 
$\kappa$ is the surface gravity. For the rotating black hole 
metrics (20) and (23), an expression for $\kappa$ is given in 
Appendix A. We can see that for the metric (20), (19) is invariant 
with respect to the rotating coordinate transformation (22).
\footnote{
 This is because $d^{D-1}x\sqrt{g_{D-1}}$ is invariant for the 
 metric (20) under the rotating coordinate transformation (22). 
 This can be seen to compare the metric (23) with the metric (20). 
 And $K-K_{0}$ is a scalar in (19).} 
Therefore the entropy calculated with the metric (23) is equal to 
that calculated with the metric (20). Hence we conclude that the 
entropy of a rotating black hole is unchanged under the rotating 
coordinate transformation (22) of its metric.

  The space-like outward-pointing normal vector on the horizon 
can be written $(0, \sqrt{g_{rr}}, 0, ...,  0)$. Hence, in (28), for 
$\omega^{\mu}$, we only need to obtain $\omega^{1}$. We use 
$0$, $1$, ..., $D-1$ to represent $\tau$, $r$, ..., $\phi_{j}$. 
For the metric (23), the only non-zero component of $g^{1\nu}$ is 
$g^{11}$. Thus, from (16) we have 
$$
  \omega^{1}=-\frac{2}{\sqrt{g}}\left( \frac{\partial\sqrt{g}}
                   {\partial r} \right)g^{rr}-
                  \frac{\partial g^{rr}}{\partial r} ~.
  \eqno{(29)}  $$
The determinant of the metric (23) is given by 
$$
  g=G_{\tau\tau}g_{rr}\det g_{ab}-
    g_{\tau\phi_{i}^{\prime}}^{2}(...)_{i} ~,
  \eqno{(30)}  $$
where $(...)_{i}$ represents terms that we need not consider  
explicitly here. The reason that these need not be included here 
is that in (29), the partial derivative of $\sqrt{g}$ is with 
respect to $r$. In (30), $g$ has two parts. We can expand 
$\sqrt{g}$ as 
$$
  \sqrt{g}=\sqrt{G_{\tau\tau}g_{rr}\det g_{ab}}
           \left[1-\frac{1}{2}\frac{g_{\tau\phi_{i}^{\prime}}^{2}
          (...)_{i}}{G_{\tau\tau}g_{rr}\det g_{ab}}+...\right] ~.
  \eqno{(31)}  $$
The second part of (30) contributes zero near the horizon, because 
the metrics $g_{\tau\phi_{i}^{\prime}}$ are squares and the metrics 
$g_{\tau\phi_{i}^{\prime}}$ are zero on the horizon, while we 
evaluate (28) on the horizon, we consider only the contribution from 
the horizon in (28). Therefore, in the following, for convenience, 
we only consider the first term in (30). This means that we consider 
the limit $r\rightarrow r_{h}$ in the integral of (28).

  Dropping the second term in (30), we obtain 
$$
  [\omega^{\mu}n_{\mu}]=-\frac{2}{\sqrt{g_{rr}}}
     \frac{\partial \ln\sqrt{G_{\tau\tau}\det g_{ab}}}
     {\partial r}-\frac{2(D-2)}{r} ~,
  \eqno{(32)}  $$
where the second term comes from the background flat spacetime.
Thus from (28), we obtain
$$
  S=-\frac{1}{4\kappa} \int d\theta \prod d\psi_{i} 
             \prod d\phi_{j}^{\prime} \frac{1}{2}
    \left[\sqrt{g_{D-1}}
      \left(-\frac{2}{\sqrt{g_{rr}}}
    \frac{\partial\ln\sqrt{g_{D-1}}}{\partial r}-\frac{2(D-2)}{r}
      \right)\right]\Bigg\vert_{r=r_{h}} ~,
  \eqno{(33)}  $$   
where $g_{D-1}$ is given by (27). The flat spacetime background 
term in (33) contributes zero, because $G_{\tau\tau}=0$ on the horizon. 
The first term inside the integral in (33) is a limit of $0\cdot\infty$. 
Because $\kappa$ is a constant on the horizon,$^{19)}$ we can move it 
inside the integral. Then, using the expression for $\kappa$ given in 
(A$~\cdot~$7), we obtain  
$$
  S=\frac{1}{4} \int d\theta \prod d\psi_{i} 
             \prod d\phi_{j}^{\prime}
     \left[\frac{1}{\sqrt{g_{rr}}}
     \frac{\partial \sqrt{G_{\tau\tau}\det g_{ab}}}
     {\partial r}\Bigg\vert_{r=r_{h}}
     \cdot \frac{\sqrt{g_{rr}}}{\partial_{r} \sqrt{G_{\tau\tau}}}
     \Bigg \vert_{r=r_{h}} \right] ~.
  \eqno{(34)}  $$
Because $\sqrt{G_{\tau\tau}}$ is zero on the horizon and $1/\kappa$ is 
finite, we finally obtain 
$$
  S=\frac{1}{4} \int d\theta \prod d\psi_{i} 
             \prod d\phi_{j}^{\prime}
            \Big[\sqrt{\det g_{ab}}\Big]\Big \vert_{r=r_{h}} ~.
  \eqno{(35)}  $$
As stated above, $g_{ab}$ denotes the metrics of the $d\theta^{2}$, 
$d\psi_{i}^{2}$, $d\theta d\psi_{i}$, $d\psi_{i}d\psi_{j}$, 
$d\phi_{i}^{\prime~2}$ and $d\phi_{i}^{\prime}d\phi_{j}^{\prime}$ 
terms in the metric (23). Hence it is seen that the above integral 
is just the area of the horizon with the metric (23). 
Then note that $g_{ab}$ here is the same as that in the metric (20), 
despite the coordinate transformation (22), i.e., 
the metrics (20) and (23) do not depend on $\phi_{i}$ and 
$\phi_{i}^{\prime}$. Then the above area of the horizon with the 
metric (23) is equal to that with the metric (20), which is given 
by (21). This means that the area of the horizon of a rotating black 
hole does not change under the rotating coordinate 
transformation (22). Thus, the Bekenstein-Hawking entropy of a 
rotating black hole with the metric (20) is given by
$$
  S=\frac{1}{4}\int d\theta \prod d\psi_{i} 
             \prod d\phi_{j}
             \Big[\sqrt{\det g_{ab}}\Big]\Big \vert_{r=r_{h}} ~,
  \eqno{(36)}  $$
which is equal to one-fourth of the area of its horizon according 
to (21). This completes the proof of the relation $S=\frac{1}{4}A$ 
for higher-dimensional rotating black holes.

\section{Conclusion}

\indent

  In this paper, we have studied the Bekenstein-Hawking entropy of 
higher-dimensional rotating black holes using the Euclidean 
path-integral method of Gibbons and Hawking. A black hole can be 
regarded as a thermodynamic system described by a grand canonical 
ensemble. With the Euclidean path-integral, grand canonical ensemble 
approach,$^{3),4)}$ Gibbons and Hawking found that the entropy of a 
black hole, not considering quantum corrections, is determined 
by the gravitational surface term. Using the gravitational surface 
term of Landau and Lifshitz,$^{10)}$ we gave a general proof 
demonstrating that the Bekenstein-Hawking entropy, not considering 
quantum corrections, is equal to one-fourth of the area of its 
horizon for general higher-dimensional rotating black holes. The 
explicit form of the area of the horizon for higher-dimensional 
rotating black holes in terms of the metrics of Myers and 
Perry$^{13)}$ was recently calculated by Jung et al.$^{20)}$

\vskip 0.8cm

\centerline{\bf Acknowledgements}

\vskip 0.2cm

  We are grateful for comments made by the referees on the 
original form of this paper. The content of the paper was 
reformed and improved following the referees' comments.

\newpage 

\centerline{\bf Appendix A} 

\centerline{ -------- {\sl Surface Gravity for Higher-Dimensional 
Rotating Black Holes} -------- }

\vskip 0.4cm

  For the metric (20), there exists the Killing field
$$
  \xi^{\mu}=\frac{\partial}{\partial\tau}+
       \sum\Omega_{i}\frac{\partial}{\partial \phi_{i}} ~,
  \eqno{({\rm A}\cdot1)}$$
where the quantities $\Omega_{i}$ are the angular velocities of 
the horizon corresponding to the coordinates $\phi_{i}$. Because 
$\xi^{\mu}$ is normal to the horizon, we have 
$$
  \xi^{\mu}\xi_{\mu}=0 
  \eqno{({\rm A}\cdot2)}$$  
on the horizon. For the metric (20), we obtain
$$
  \xi^{\mu}\xi_{\mu}=g_{\tau\tau}+g_{\phi_{i}\phi_{i}}\Omega_{i}^{2}+
         \sum_{i<j}2g_{\phi_{i}\phi_{j}}\Omega_{i}\Omega_{j}+
         2g_{\tau\phi_{i}}\Omega_{i} ~.
  \eqno{({\rm A}\cdot3)}$$
Next, we define 
$$
    G_{\tau\tau}=g_{\tau\tau}+g_{\phi_{i}\phi_{i}}\Omega_{i}^{2}+
         \sum_{i<j}2g_{\phi_{i}\phi_{j}}\Omega_{i}\Omega_{j}+
         2g_{\tau\phi_{i}}\Omega_{i} ~.
  \eqno{({\rm A}\cdot4)}$$
Therefore, we have
$$
    G_{\tau\tau}\big\vert_{r=r_{h}}=0 ~.
  \eqno{({\rm A}\cdot5)}$$
The surface gravity $\kappa(r_{h})$ is constant 
on the horizon.$^{19)}$ Writing $\xi^{\mu}\xi_{\mu}=-\lambda^{2}$, 
the surface gravity can be determined from the equation$^{21)}$ 
$$
  \nabla^{\mu}(\lambda^{2})\nabla_{\mu}(\lambda^{2})
   =-4\kappa^{2}\lambda^{2} ~.
  \eqno{({\rm A}\cdot6)}$$
Thus, for the metric (20), we obtain 
$$
  \kappa(r_{h})=\lim_{r\rightarrow r_{h}}
     \frac{\partial_{r}\sqrt{G_{\tau\tau}}}{\sqrt{g_{rr}}} ~.
  \eqno{({\rm A}\cdot7)}$$
The partial derivative here is taken before the limit, because 
$G_{\tau\tau}$ is zero on the horizon. From (A$~\cdot~$6), we see 
that $\kappa$ is a scalar. Hence, the surface gravity of a rotating 
black hole is invariant under general coordinate transformations.

\vskip 0.8cm

\centerline{\bf References}

\vskip 0.4cm 

1) J. D. Bekenstein, Lett. Nuo. Cim. {\bf 4} (1972), 737;
   Phys. Rev. D {\bf 7} (1973), 2333; 

   ~~~ Phys. Rev. D {\bf 9} (1974), 3292.

2) S. W. Hawking, Nature (London) {\bf 248} (1974), 30;
   Commun. Math. Phys. {\bf 43} (1975),

   ~~~ 199.

3) G. W. Gibbons and S. W. Hawking, Phys. Rev. D {\bf 15} (1977), 2752. 

   ~~~ G. W. Gibbons, S. W. Hawking and M. J. Perry, 
       Nucl. Phys. B {\bf 138} (1978), 141.

4) S. W. Hawking, in {\sl General Relativity: An Einstein Centenary
     Survey}, ed. S. W. 

     ~~~ Hawking and W. Israel (Cambridge University 
     Press, Cambridge, England, 1979),  

     ~~~ Chap. 15.

5) G. W. Gibbons and M. J. Perry, Proc. R. Soc. Lond. A {\bf 358}
   (1978), 467.

6) S. W. Hawking and G. F. R. Ellis, {\sl The Large Scale Structure of
   Spacetime} (Camb

    ~~~ ridge Univ. Press, Cambridge, England, 1973).

7) R. Kallosh, T. Ortin and A. Peet, Phys. Rev. D {\bf 47} (1993), 5400, 
   hep-th/9211015. 

8) S. Liberati and G. Pollifrone, Phys. Rev. D {\bf 56} (1997), 6458, 
   hep-th/9708014.

9) B. S. DeWitt, Phys. Rev. {\bf 160} (1967), 1113.

10) L. D. Landau and E. M. Lifshitz, {\sl The Classical Theory of
    Fields} (Pergamon, 

    ~~~~ Oxford, 1975).

11) H. W. Braden, J. D. Brown, B. F. Whiting and J. W. York, Jr., 
    Phys. Rev. D {\bf 42}

    ~~~~ (1990), 3376. 

12) A. Chamblin, R. Emparan, C. V. Johnson and R. C. Myers, 
    Phys. Rev. D {\bf 60} (1999), 

    ~~~~ 064018, hep-th/9902170. 

13) R. C. Myers and M. J. Perry, Ann. of Phys. {\bf 172} (1986), 304.

14) M. Cveti\u{c} and D. Youm, Nucl. Phys. B {\bf 477} (1996), 449, 
    hep-th/9605051. 

15) C. J. Hunter, Phys. Rev. D {\bf 59} (1999), 024009, gr-qc/9807010.

16) S. W. Hawking, C. J. Hunter and M. M. Taylor-Robinson, 
    Phys. Rev. D {\bf 59} 

    ~~~~ (1999), 064005, hep-th/9811056. 

17) S. W. Hawking, G. T. Horowitz and S. F. Ross, Phys. Rev. 
    D {\bf 51} (1995), 4302, 

    ~~~~ gr-qc/9409013.

18) C. Teitelboim, Phys. Rev. D {\bf 51} (1995), 4315, hep-th/9410103. 

19) J. M. Bardeen, B. Carter and S. W. Hawking, 
    Commun. Math. Phys. {\bf 31} (1973),   

    ~~~~ 161. 

20) E. Jung, S. H. Kim and D. K. Park, Phys. Lett. B {\bf 619} 
    (2005), 347, hep-th/0504139. 

21) R. M. Wald, {\sl General Relativity} (The University of Chicago 
    Press, 1984).

\end{document}